# NOTES ON FREQUENCIES OF OCCURRENCE AND ENERGETICS OF THE SOLAR-TYPE STELLAR FLARES

**R.E. Gershberg (Crimean Astrophysical Observatory, Nauchny Crimea Russia)**

Abstract

On the basis of the 30-year ago ground-based photometry and the recent Kepler space experiment there have been considered frequencies of occurrence and energetics of the solar-type stellar flares. It was concluded that frequencies of occurrence of such flares are proportional to sizes of stellar surfaces, and estimates of maximum flare radiation from the results of the ground-based photometry and space observations practically coincide.

## 1. INTRODUCTION

As is known, in 1958, when distinguishing flare stars into a separate type of variable objects, the red dwarf M6e L 726-8 B = UV Cet was chosen as a typical representative. And red dwarfs predominated in lists of such variables when their number achieved hundreds – see Table containing spectral distributions of flare stars from the Crimean catalogues by Gershberg et al. (1999) and (2011), designated in Table as GKL99 and GTSh10, respectively. In the second and third lines of Table the characteristic luminosities and sizes of such stars are given. On sampling from many tens of thousands M stars in the frame of the Sloan project (SDSS), the portion of dwarfs with Hα emission, that is a characteristic of flare stars, increases in the range of M0-M9 from 1 to 80% (West et al., 2011).

*Table*

| Spectra | F | G0-G9 | K0-K3 | K4-K8 | M0-M3 | M4-M8 |
|---|---|---|---|---|---|---|
| $<L_*/L_\odot>$ | 3 | 0.9 | 0.4 | 0.2 | 0.05 | 0.006 |
| $<R_*/R_\odot>$ | 1.2 | 1.0 | 0.8 | 0.7 | 0.5 | 0.2 |
| Number of the UV Cet-type stars in GKL99 |  | 10 | 19 | 25 | 146 | 212 |
| Number of the UV Cet-type stars in GTSh10 | 10 | 13 | 23 | 66 | 224 | 230 |

After detecting radio emission from flares of the UV Cet-type stars (Lovell, 1964) and recording high time-resolution spectra of such flares (Gershberg and Chugajnov, 1966, 1967; Kunkel, 1967) Gershberg and Pikel'ner (1972) put forward an idea on the physical identity of activity in the red dwarf stars and the Sun, and Mullan (1975) involved the concept of stellar magnetism as a basis of this activity. The essential distinctions between global parameters of these stars and the Sun caused the cautious attitude to this idea, but the rapid development of the all-wavelength astrophysics in the 70-80s revealed in these stars in addition to already known sporadic optical and radio flares and cool photospheric spots almost all other phenomena of the solar activity – chromospheric active regions, thermal X-ray and non-thermal radio emission of coronae, prominence-like structures, long-term variations of photospheric radiation and various atmospheric layers similar to the solar 11-year and secular activity cycles, differential rotation at various latitudes. Therefore, today the concept of unified consideration of solar activity and activity of the red dwarf stars is generally accepted.

As numerous observations have shown, there exist systematic distinctions between average spectral types and, consequently, luminosities of dwarf stars with those or other manifestations of this activity: flares and quiet chromospheres are the most accessible for observations and are studied in the coldest M stars; among the spotted dwarfs there are mostly early M and late K stars; long-term cycles of activity are detected mainly in early K and late G dwarfs. The observational selection is clearly shown in this regularity: flares and chromospheric emission are the most noticeable against the weak photospheric radiation of the coldest dwarf stars; for the photometric detection of spots, i.e. for recording small deformations of the light curve of a star, the level of its normal brightness should be defined with high accuracy for what a brighter photosphere is required, and in radiation of such a photosphere the sporadic outbursts in brightness caused by flares will be lost. Finally, the long-term cycles of activity are detected mainly based on high-resolution spectrometric observations which are possible only in studying brighter stars. And X-ray and radio emission of coronae, for which the photospheric radiation is not an obstacle at all, are recorded from the early F to late L dwarfs. The question as to what degree an observational selection masks real dependence of the general level of various manifestations of stellar activity on their luminosity, mass and age, demands, in each case, a special analysis of observational data.

In these notes we discuss results of the ground-based and space photometric observations of flare stars concerning frequencies of occurrence and maximum energy of flares.

## 2. FREQUENCIES OF OCCURRENCE OF THE OBSERVABLE FLARES

Analyzing photoelectric observations of 8 flare stars executed by Moffett (1974), Mirzoyan (1981) has found out a growth of average frequency of recording flares with increase in absolute magnitude. Considering this question thoroughly based on more complete data, between these average frequencies and absolute magnitudes of stars the confident positive correlation has been detected:

$$r(\lg \nu, M_B) = 0.81 \pm 0.1 \qquad (1)$$

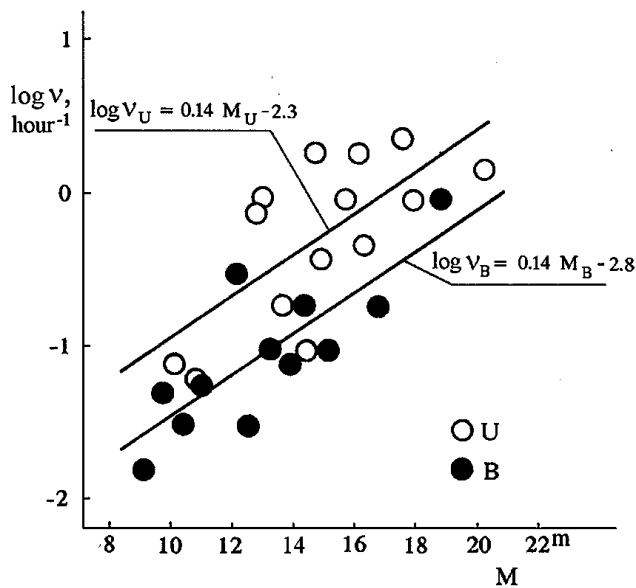

Fig. 1. Dependences of average frequencies of occurrence of the recorded flares on absolute luminosities of flare stars (Gershberg, 1985).

In Fig. 1 the comparison of these magnitudes and analogous magnitudes taken from observations of flare stars in the U band is given (Gershberg, 1985). The equations of direct linear regressions are as follows:

$$\lg \nu_U = (0.14 \pm 0.04) M_U - 2.3 \pm 0.6$$

$$\lg \nu_B = (0.14 \pm 0.03) M_B - 2.8 \pm 0.4 \quad (2)$$

So, the statistical dependence of average frequencies of occurrence of all flares observed in flare stars on absolute luminosities can be presented via correlation

$$\nu \propto L^{-0.35 \pm 0.09} \quad (3)$$

The reality of the observable dependence is beyond doubt, and the notable disorder of points in Fig. 1 and corresponding remarkable probable errors of numerical coefficients in (2) can be caused by heterogeneity of the initial observational data and physical heterogeneity – for example, in age – of the considered flare stars. Finally, the scattering of points in Fig. 1 can be accounted for by the circumstance that flare stars are often components of binary systems which can contain stars of different brightness and different level of flare activity; therefore, there should be a dispersion in observable dependence of $\lg\nu$ on M even at an existence of functional relation between these parameters for individual flare stars.

The decrease in average frequency of occurrence of the recorded flares with increase in absolute luminosity is naturally to be associated with observational selection: in brighter stars the flare-detection threshold is higher. The detailed statistical analysis of the observable spectra of flares shows that at the power form of these spectra the effect of observational selection should lead to a correlation

$$\nu \propto L^{-0.96 \pm 0.13} \quad (4)$$

(Gershberg, 1985). The significant distinction between (3) and (4) means that at the transition from absolutely weak to absolutely brighter flare stars the observable frequency of flares decreases not so quickly as it should be only due to increase of a flare-detection threshold. This effect is likely to be associated with an increase in surface area of brighter stars and existence of greater number of flare-producing active regions. Actually, taking into account Pettersen's data (1976, 1980) on absolute luminosities and sizes of single flare stars it is possible to obtain a statistical correlation between these values in the form

$$L \propto R^{5.0 \pm 0.06} \quad (5)$$

Then for the factor, correlations (3) and (4) differ in, we get

$$L^{0.6 \pm 0.2} \propto R^{3.0 \pm 1.4} \propto 4\pi R^2 \cdot R^{1.0 \pm 1.4} \quad (6)$$

The last co-factor in (6) $R^{1.0 \pm 1.4}$ is too undefined for any conclusions about a level of "specific" flare activity from the unit of the flare star's surface; but the influence of a size of the stellar surface on the frequency of the recorded flares is highly probable.

Thus, the dependence of average frequency of occurrence of the recorded flares on the absolute luminosity of a star is defined by the dependence of flare-detection threshold on the stellar luminosity as well as by the real distinctions between average frequencies of flares from the stars of different luminosity. Moreover – contrary to the widespread opinion – the true average frequency of flare occurrence is higher in brighter stars. The noted erroneous opinion appeared as a result of the fact that detected from observations event of higher frequency of the recorded flares from the low-luminosity stars has not been considered from the viewpoint of inevitable effects of observational selection. It should be noted, finally, that since statistical researches lead to the conclusion that both true frequencies of flare occurrence and their maximum luminosities in brighter flare stars are higher than in less bright, so red dwarf stars are the most pronounced carriers of the considered type of activity not because such activity in these objects is the most developed, but only thanks to that it is the most noticeable in such cold low-luminosity stars.

Here it is appropriate to remind the recent result by Balona (2015) who based on observations with the Kepler spacecraft concluded that approximately 5% of K-M dwarfs flare whereas about 1.5% of stars of each from G, F and A types reveal flare activity as well. Moreover, stars of these middle spectral types definitely burst themselves, but not their weak satellites, as it was supposed earlier. In other words, Balona has not detected an abrupt decline in the observed flare activity in the range of these middle spectral types which might be expected because of rather fast decreasing thickness of the convection zone and, hence, weakening of a dynamo generating the magnetic field. And decreasing from 5% of flaring K-M stars to 1.5% of G-F-A active stars he referred to decreasing contrast of flares on the background of photospheres with the growing temperature.

### 3. FLARES WITH MAXIMUM ENERGY

As is known, Maehara et al. (2012) have analyzed 365 strong flares with energy more than $10^{33}$ erg, recorded from 148 G dwarfs during observations of 83000 stars in terms of the Kepler experiment over 120 days in 2009. The typical duration of such flares was several hours, amplitude – 0.1-1% of the stellar bolometric luminosity; when making energy estimations it was supposed that flares radiate as absolutely black bodies at the temperature of 10000 K. The estimates of bolometric luminosity of such flares were from $9·10^{29}$ to $4·10^{32}$ erg/sec and their total energy – from $10^{33}$ to $10^{36}$ erg; the uncertainty of these estimates reaches 60%. The detected by Maehara et al. quasi-periodicity evidences about greater number of the spotted regions than in the Sun. The maximum energy of flares does not correlate with periods of rotation, but powerful flares occur more often on stars with fast rotation. Proceeding from the derived estimates, Maehara et al. have estimated that such powerful flares should occur from a G star once per 350 years. In slowly rotating stars as the Sun such flares should occur once per 800 years, and flares with energy $10^{35}$ erg – once per 5000 years. And if it is true, the Earth has already undergone about one million (!) such tremendous solar flares without (?) any visible influence on the terrestrial life.

Later, Shibayama et al. (2013) repeated researches of Maehara et al. (2012) on the more complete data when during 500-day monitoring with Kepler 1549 superflares have been recorded from 279 G stars. According to these data, the energy spectrum of superflares has a power form $dN/dE \propto E^{-\alpha}$ with $\alpha \sim 2$ and in such slowly rotating stars flares with energy $10^{34}$– $10^{35}$ erg occur once in 800–5000 years; but in some G stars for 500 days 57 superflares have been recorded; evidently, there exist huge spots and their groups. Among stars with flares with energies in the range of $10^{33}$–$10^{36}$ erg Wihmann et al. (2014) have found out very young fast rotators.

Finally, Candelaresi et al. (2014) have considered 117661 G-K-M star observed with Kepler, and 1690 flares with energy more than $10^{34}$ erg have been detected from 380 of them. They used this data to ascertain influence of effective temperature and rotation velocity on the occurrence of such flares and their energy, and found out that with growth of effective temperature the occurrence of superflares decreases. For slowly rotating stars they have found out the square-law growth of occurrence of such flares to some critical limit after which it linearly decreases. Among quickly rotating stars the increasing portion of stars is characterized by higher spottedness that leads to greater occurrence of flares.

Let's compare results of space researches and ground-based photometry.

The paper by Gershberg, Mogilevsky and Obridko (1987) presented and involved the analysis of energy spectra of flares in individual stars of the solar vicinity and group spectra of stellar flares in the Pleiades and Orion stellar clusters. The technique of photographic observations of simultaneously several hundreds of flare stars in clusters allows recording very rare events, and the ergodic principle – to unite flares of close in brightness flaring stars in a cluster for calculating group spectra of flares – see Fig. 2. For decreasing scatter in data the cumulative frequencies are placed along the abscissa axis.

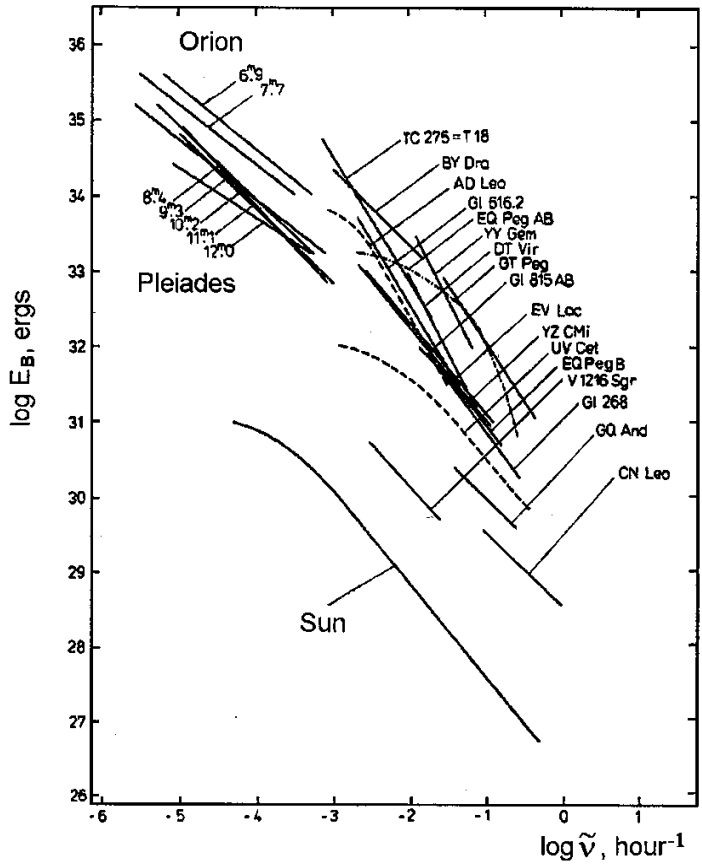

Fig. 2. Energy spectra of stellar flares in the solar vicinity, spectra of the Pleiades and Orion stellar clusters and solar flares (Gershberg et al., 1987)

As it follows from Fig. 2 where energy spectra of flares are given in double logarithmic scale, the maximum value of integral energy of flares in the B band – the $E_B$ value – is not less than $3 \cdot 10^{35}$ erg. Supposing that the total optical energy of flares is close enough to their bolometric energy, and using a correlation between flare energies from the above paper

$$E_{опт} = 4.2\, E_B \qquad (7)$$

we get $10^{36}$ erg for the maximum value of energy, almost coinciding with the Kepler's estimates.

Thanks to M.M. Katsova and M.A. Livshits for useful discussions and Ya.V.Poklad for improving my English.

REFERENCES

*Balona L.A., 2015 – Monthly Notices RAS V.447 is.3 P.271.*
*West A.A., Morgan D.P., Bochanski J.J., et al.), 2011 — Astron.J. V.141 P.97.*
*Wichmann R., Fuhrmeister B., Wolter U., Nagel E.), 2014 – arXive: 1406.0612v1 3 Jun 2014.*
*Gershberg R.E., 1985 — Astrofizika V.22, P.531.*
*Gershberg R.E. and Chugainov P.F., 1966 — Soviet Astron. V.43 P.1168.*
*Gershberg R.E. and Chugainov P.F., 1967 — Soviet Astron. V.44 P.260.*
*Gershberg R.E. and Pikel'ner S.B. 1972 — Comments Astrophys. Space Physics V.4 P.113.*
*Gershberg R.E., Mogilevsky E.I., Obridko V.N., 1987 — Cinematics and physics of celestial bodies, V.3 №5 P.3.*
*Gershberg R.E., Katsova M.M., Lovkaya M.N., Terebizh A.V., Shakhovskaya N.I., 1999 — Astron. Astrophys. Suppl.Ser. V.139 P.555.*
*Gershberg R.E., Terebizh A.V., Shlyapnikov A.A., 2011 — Bul. Crimean astrophys.Obs. V.107 Is.1, P.18.*
*Candelaresi S., Hillier A., Maehara H., Brandenburg A., Shibata K., 2014 – Astrophys.J. V.792 P.97.*

*Kunkel W., 1967 — «An optical study of stellar flares». Theses. University of Texas.*
*Lovell B., 1964 — Observatory V.84 P.191.*
*Maehara H., Shibayama T., Notsu S., Notsu Y., et al., 2012 — Nature V.485 P.478.*
*Mullan D.J., 1975 — Astron.Astrophys. V.40 P.41.*
*Mirzoyan L.V., 1981 — «Nonstationarity and evolution of stars». Armenian AS publ., Erevan 380P.*
*Moffett T.J., 1974 — Astrophys.J.Suppl.Ser. V.29 P.1.*
*Pettersen B.R., 1976 — Theoret.Astrophys.Institute Report N 46. Blindern-Oslo.*
*Pettersen B.R., 1980 — Astron.Astrophys. V.82 P.53.*
*Shibayama T., Maehara H., Notsu S., Notsu Y., et al., 2013 — Astrophys.J. Suppl.Ser. V.209 P.5.*